# Second-harmonic generation using $\bar{4}$-quasi-phasematching in a GaAs microdisk cavity

Paulina S. Kuo and Glenn S. Solomon

*Joint Quantum Institute, National Institute of Standards and Technology & University of Maryland, Gaithersburg, MD 20899*

Corresponding emails: pkuo@nist.gov; solomon@nist.gov

**The $\bar{4}$ crystal symmetry in materials such as GaAs can enable quasi-phasematching[1,2] for efficient frequency conversion without poling, twinning or other engineered domain inversions[3–5]. $\bar{4}$ symmetry means that a 90° rotation is equivalent to a crystallographic inversion. Therefore, propagation geometries where light circulates about the $\bar{4}$ axis produce effective domain inversions, useful for quasi-phasematching. Microdisk optical cavities also offer resonance field-enhancement and excellent spatial overlap, resulting in highly efficient frequency conversion in micrometer-scale volumes. These devices can be integrated in photonic circuits as compact sources of radiation or entangled photons[6,7]. Efficient second-order frequency conversion is a new functionality for nonlinear semiconductor microdisk resonators, which have been previously explored for all-optical circuits as switches, signal routers and optical logic gates[8,9]. Arrays of microresonators can produce slow light[10–12] and robust optical delay lines[13]. Here, we present the first experimental observation of efficient second-harmonic generation in a microdisk cavity utilizing $\bar{4}$-quasi-phasematching.**

Efficient optical frequency conversion requires the phases of the driving polarization and generated electric field be matched or properly compensated[1,2,14]. Consider a sum-frequency process where $\omega_3 = \omega_1 + \omega_2$. The generated electric field at frequency $\omega_3$ propagates as $E_3 \propto \exp(-ik_3 z)$ while the nonlinear driving polarization propagates as $P_3^{NL} \propto \chi^{(2)} E_1 E_2 \propto \exp[-i(k_1 +$ $k_2)z$], where $k_i$ is the wavevector at frequency $\omega_i$ and $\chi^{(2)}$ is the second-order nonlinear coefficient. Perfect phasematching is achieved when $k_3 = k_1 + k_2$ and can be obtained by, for instance, mixing orthogonally polarized waves in a birefringent crystal[14]. Efficient conversion can also be attained using quasi-phasematching (QPM)[1,2], where $\chi^{(2)}$ is modulated at period $\Lambda = 2\pi j/|k_3 - k_1 - k_2|$, ($j$ is an integer). This modulation can be produce by periodic domain inversions where $\chi^{(2)}$ changes sign[1,2], by on-off modulation[15], or by modulation in the magnitude of $\chi^{(2)}$ [16]. Materials such as GaAs that possess $\bar{4}$ crystal symmetry can exhibit an effective $\chi^{(2)}$ modulation when the fields propagate in curved geometries (such as in microrings or microdisks)[3–5]. A 90° rotation about the $\bar{4}$ axis is the same as a crystallographic inversion, and hence fields propagating around the $\bar{4}$ axis in a uniform GaAs microdisk effectively encounter four domain inversions per round trip. The $\bar{4}$ crystal symmetry allows QPM to be achieved without externally produced domain inversions.

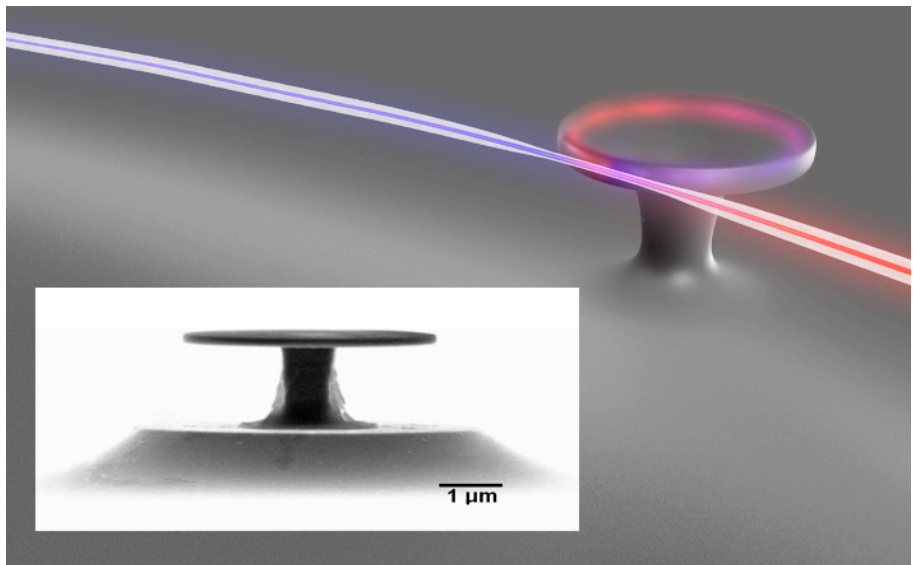

**Figure 1 Schematic of second-harmonic generation in a fiber-taper-coupled GaAs microdisk.** Input light at the fundamental wavelength (represented by red light) is converted inside the microdisk to second-harmonic light (represented by blue light). The inset shows a scanning electron micrograph of a fabricated device.

A <001>-surface-normal GaAs microdisk with perfect cylindrical symmetry supports both TM (electric field orthogonal to the disk plane) and TE (electric field in the plane of the disk) polarizations. From the $\chi^{(2)}$ tensor for GaAs and other $\bar{4}3m$ point-group materials, one of the three interacting fields must be TM-polarized while the other two are TE-polarized. Consider second-harmonic generation (SHG) in a microdisk where the fundamental, $\omega_f$, is converted to the second harmonic (SH), $\omega_{SH} = 2\omega_f$ (Fig. 1). The electric field at each frequency ($\omega_i$, $i = f$ or $SH$) is

$$E_i = A_i(\theta)e^{i(\omega_i t - m_i \theta)}, \qquad (1)$$

where $A_i(\theta)$ is the slowly varying amplitude, and $m_i$ is the azimuthal number, which is an integer for a resonant mode. From Maxwell's equations and the $\chi^{(2)}$ tensor projections in the circular propagation geometry, the change in the SH amplitude with propagation angle θ is[3,17]

$$\frac{dA_{SH}}{d\theta} = A_f^2 \left( K_+ e^{i(\Delta m + 2)\theta} + K_- e^{i(\Delta m - 2)\theta} \right), \qquad (2)$$

where $\Delta m = m_{SH} - 2m_f$, and $K_+$ and $K_-$ are the SHG coefficients calculated from the mode-overlap integrals[17]. SHG is largest when the QPM condition is satisfied: $\Delta m$ = +2 or -2. This condition is equivalent to requiring the microdisk circumference to essentially equal $4\pi/|k_{SH} - 2k_f|$[4,5].

$\bar{4}$-quasi-phasematched, second-order nonlinear mixing in GaAs whispering-gallery-mode (WGM) microdisks differs from nonlinear mixing in other microresonators. Efficient SHG has been observed in a WGM resonator made of periodically poled LiNbO$_3$[18] where QPM was achieved using engineered domain inversions. Birefringent phasematching was utilized to demonstrate efficient mixing in a uniform LiNbO$_3$ WGM resonator [19]. Resonantly enhanced SHG has been observed in a GaP photonic crystal cavity[20]. Also, frequency conversion using four-wave mixing ($\chi^{(3)}$) has been shown in GaAs microring resonators[21]. $\chi^{(2)}$-based frequency conversion in GaAs WGM resonators using $\bar{4}$-QPM is potentially more efficient than these other demonstrations since the nonlinear susceptibility of GaAs is larger than that of LiNbO$_3$ or GaP[22]. In addition, LiNbO$_3$ resonators are formed by mechanical polishing rather than microfabrication processing, and may be difficult to integrate in photonic circuits.

The most efficient SHG is obtained when both the fundamental and SH waves are resonant with the microdisk ($\lambda_f = 2\lambda_{SH}$, where $\lambda_i$ are resonance wavelengths), and the $\bar{4}$-quasi-phasematching condition is satisfied ($\Delta m = \pm 2$). However, frequency conversion can still be obtained if one of the waves is off-resonance with the microdisk ($\lambda_f \neq 2\lambda_{SH}$), which leads to partially-doubly-resonant or singly-resonant SHG. The generated second-harmonic spectrum, $P_{SH}^{out}$, is proportional to[17]

$$P_{SH}^{out} \propto \left(P_f^{circ}\right)^2 P_{SH}^{circ} \left|\tilde{K}^{NL}\right|^2, \qquad (3)$$

where $P_f^{circ}$ and $P_{SH}^{circ}$ are the circulating-power spectra of the microdisk cavity in the fundamental and SH wavelength ranges, and $|\tilde{K}^{NL}|^2$ is the nonlinear gain from one round trip. The circulating-power spectra are typically much narrower than the nonlinear gain spectrum. Equation (3) implies that partially-doubly-resonant SHG is more efficient when only the fundamental wave is fully on-resonance than when only the SH wave is fully on-resonance.

The amount of SH conversion depends on the detuning of the resonances ($\lambda_f - 2\lambda_{SH}$) compared to their linewidths. The resonance linewidths depend on the coupling ($Q_i^c$) and intrinsic ($Q_i^0$) quality factors of the microdisk, which are related to the total quality factors by $1/Q_i^{tot} = 1/Q_i^c + 1/Q_i^0$. $Q_i^0$ are typically fixed by fabrication while $Q_i^c$ can be varied by adjusting the coupling rate into the microdisk. When the resonances are aligned (doubly-resonant SHG is achieved; $\lambda_f - 2\lambda_{SH}=0$), maximum conversion is obtained at critical coupling ($Q_i^c = Q_i^0$ at both wavelengths). When the resonances are not aligned ($|\lambda_f - 2\lambda_{SH}| \neq 0$), higher SH conversion may be produced when the cavity is over-coupled ($Q_i^c < Q_i^0$) where increased coupling broadens the resonance linewidths and produces better spectral overlap, as illustrated in Fig. 2. Using theory presented in Ref. 17, we calculated SH conversion, $\eta$, for different resonance detunings, $|\lambda_f - 2\lambda_{SH}|$, as $Q_i^c$ is varied relative to $Q_i^0$. In Fig. 2a, $Q_f^c$ is varied relative to $Q_f^0$ while we fix $Q_{SH}^c = Q_{SH}^0$. In addition to the peak conversion at $Q_f^c = Q_f^0$, a second maximum is obtained at $Q_f^c / Q_f^0 < 1$ when the half width at half maximum (HWHM) of the fundamental resonance equals the detuning. Figure 2b shows the SH conversion when $Q_{SH}^c$ is varied relative to $Q_{SH}^0$ while we fix $Q_f^c = Q_f^0$. Only when $|\lambda_f - 2\lambda_{SH}| = 0$ nm is the SH conversion maximized at $Q_{SH}^c = Q_{SH}^0$. With non-zero $|\lambda_f - 2\lambda_{SH}|$, the conversion efficiency is largest at lower $Q_{SH}^c$ where the SH resonance is broadened by coupling. In general, increased detunings lower the peak SH conversion (see Fig. 2 insets), but reducing the $Q_i^c/Q_i^0$ ratio can help mitigate the drop. Over-coupling broadens the resonance linewidths, which can lead to better spectral overlap and improved conversion.

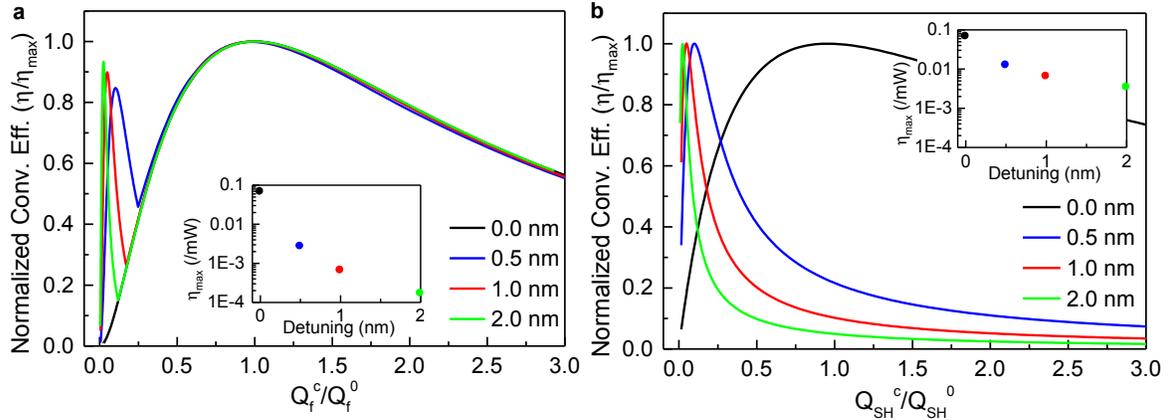

**Figure 2 Dependence of SH conversion efficiency, $\eta$, on resonance detuning and quality-factor ratio**. $\eta$ is normalized to its peak value and shown as a function of the ratio of coupling to intrinsic quality factors ($Q_i^c / Q_i^0$) for several detunings ($|\lambda_f - 2\lambda_{SH}|$ = 0, 0.5, 1.0, and 2.0 nm). It is calculated for doubling $\lambda_f \approx 1990$ nm in a 2.6-μm-radius, 160-nm-thick GaAs microdisk where $\Delta m = -2$. The intrinsic quality factors are $Q_f^0 = Q_{SH}^0 = 20000$. In **a**, critical coupling is assumed for the SH resonance ($Q_{SH}^c = Q_{SH}^0$) while $Q_f^c$ is varied relative to $Q_f^0$, while in **b,** critical coupling is assumed for the fundamental resonance ($Q_f^c = Q_f^0$) while $Q_{SH}^c$ is varied relative to $Q_{SH}^0$. When $|\lambda_f - 2\lambda_{SH}| \neq 0$, optimum SH conversion may be obtained by lowering $Q_i^c$ and over-coupling until the HWHM equals the detuning. The insets plot the value of maximum SH conversion efficiency ($\eta_{max}$) for the different detunings. Larger detunings lead to lower conversion efficiency, but over-coupling to broaden linewidths can improve conversion.

Experimentally, GaAs microdisks were probed using the setup sketched in Fig. 3. Microdisk resonances are characterized by three integers $(m_i, p_i, q_i)$ that count the number of azimuthal, radial and vertical antinodes, respectively. Using a tapered optical fiber to launch and collect light from microdisk[23] (Fig. 1), we measured the transmission spectra of the microdisks in the fundamental ($\lambda_f$=1900-2015 nm) and SH ($\lambda_{SH}$=950-1007 nm) wavelength ranges and identified ($m_i, p_i$) values of the resonances ($q_f = q_{SH} = 1$ for the thin microdisks used here). The resonance wavelengths and free-spectral ranges of different radial mode families were compared to predictions from finite-element modeling to determine ($m_i, p_i$). The microdisk thickness and radius were used as fitting parameters in the modeling, and the resulting fitted sizes were consistent with observations by scanning electron microscopy. We characterized $Q_i^0$ and $Q_i^c$ by observing the changes in the transmission spectra as the gap between the fiber taper and microdisk was varied. The microdisks were then pumped with the fundamental beam, and generated SH light was detected by a silicon detector with a dichroic filter to reject residual pump light.

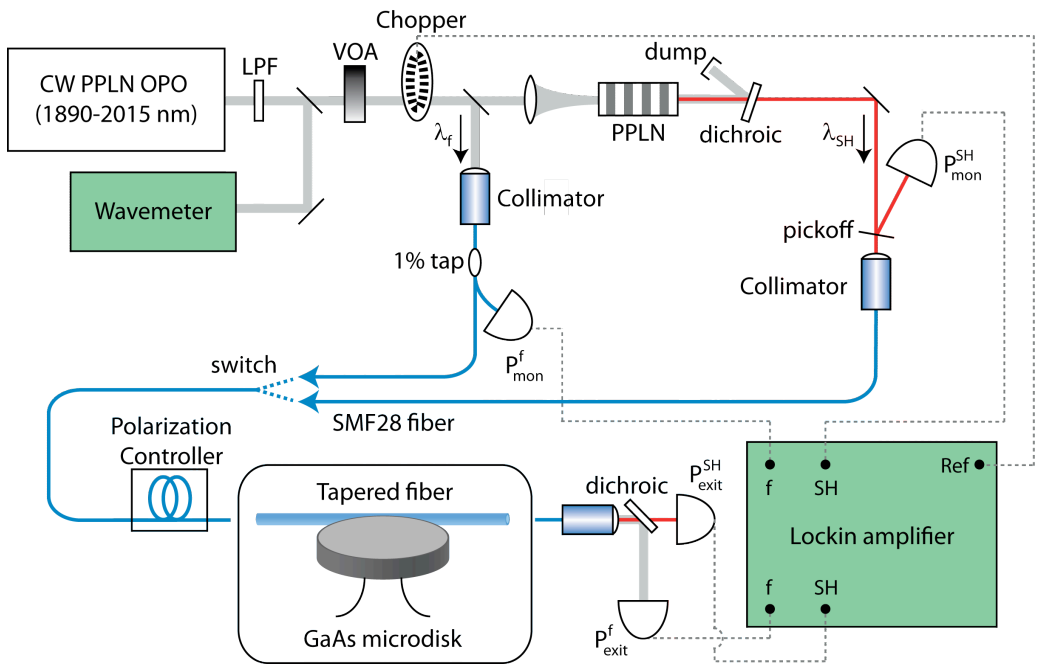

**Figure 3 Diagram of the experimental setup.** Two fiber-coupled beams at the fundamental and SH wavelengths were produced using a continuous-wave (CW), periodically poled LiNbO$_3$ (PPLN) optical parametric oscillator (OPO) and a PPLN doubling crystal. The beams were alternately coupled into the GaAs microdisk using a tapered fiber. The monitor ($P_{mon}^f, P_{mon}^{SH}$) and exit ($P_{exit}^f, P_{exit}^{SH}$) powers were recorded by a lockin amplifier. LPF, 1850-nm long-pass filter; VOA, variable optical attenuator; dichroic filter reflects the fundamental and transmits the second-harmonic.

We fabricated GaAs microdisks with 2.6-µm radius and 160-nm thickness using molecular-beam epitaxy, photolithography and wet chemical etching (see Methods for details). Figure 1 (inset) shows a scanning electron micrograph of the GaAs microdisk cavity. All fabricated microdisks showed non-zero detuning of the fundamental and SH resonances. Figure 4 shows the fundamental and SH transmission spectra for the GaAs microdisk with the smallest detuningmeasured with the fiber taper in contact with the microdisk. We identified the TE-polarized, fundamental resonance centered at 1985.38 nm as ($m_f$=13, $p_{SH}$=1). At the SH wavelength, we observed two TM-polarized resonances in the vicinity of 992 nm. By probing these resonances with various taper-disk separations and fiber taper sizes, we determined the resonance at 991.6 nm is ($m_{SH}$=24, $p_{SH}$=2) and the resonance at 992.25 nm is ($m_{SH}$=21, $p_{SH}$=3). We also observed the ($m_{SH}$=28, $p_{SH}$=1) resonance for this microdisk at 997 nm. Nonlinear optical mixing using the $m_f$=13 and $m_{SH}$=21 resonances does not occur since there is no phasematching ($\Delta m$ = -5)[17], and the ($m_{SH}$=28, $p_{SH}$=1) resonance is too far detuned for efficient SHG. We concluded that SHG in the microdisk involves the ($m_f$=13, $p_f$=1) fundamental resonance at 1985.38 nm and the ($m_{SH}$=24, $p_{SH}$=2) SH resonance at 991.6 nm ($\Delta m$ = -2) with detuning $|\lambda_f - 2\lambda_{SH}|$=2.2 nm.

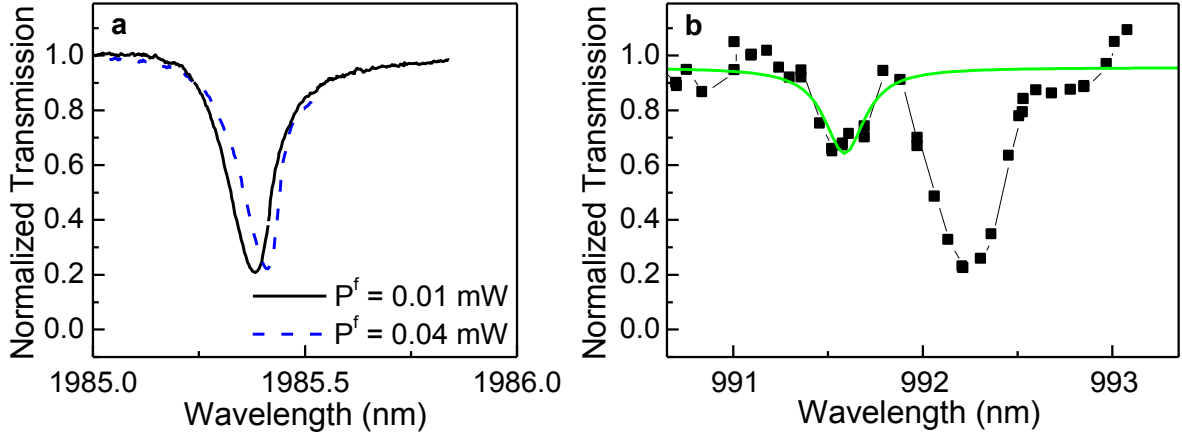

**Figure 4 Measured transmission spectra of the microdisk cavity. a,** TE-polarized transmission spectra of the ($m_f$=13, $p_f$=1) resonance near 1985 nm at two power levels. **b,** TM-polarized, transmission spectrum (black) near 990 nm along with fit (green) to the ($m_{SH}$=24, $p_{SH}$=2) resonance.

Figure 5a plots the observed SH conversion efficiency ($P^{SH}/P^f$) as a function of the fundamental pump wavelength. All powers are given for the point inside the fiber taper where the fiber touches the GaAs microdisk and are calculated based on the measured taper losses and observed external powers. SH conversion is maximized when the fundamental is on-resonance while the SH is off-resonance, as seen by Fig. 5a top inset. Figures 5b and 5c plot the generated SH power and the SHG conversion efficiency, respectively, for $\lambda_f$ = 1985.38 nm. At low pump power, the logarithmic plot of generate SH power as a function of fundamental power (Fig. 5b) has a slope of 2.15±0.10, confirming quadratic growth of the second-harmonic. We found the normalized conversion efficiency, $\eta$, at low power was (5.2 ± 0.4)×10$^{-5}$/mW, while at higher power, the conversion efficiency rolled off. From Fig. 4, we observed $Q_f^{tot}$=16000 and $Q_{SH}^{tot}$=4000. We estimated the intrinsic quality factors from transmission spectra taken with the taper several hundred nanometers separated from the microdisk (where $1/Q_i^c \approx 0$), and found $Q_f^0$=33000 and $Q_{SH}^0$=9000. With these quality factors and resonance detuning $|\lambda_f - 2\lambda_{SH}|$ =

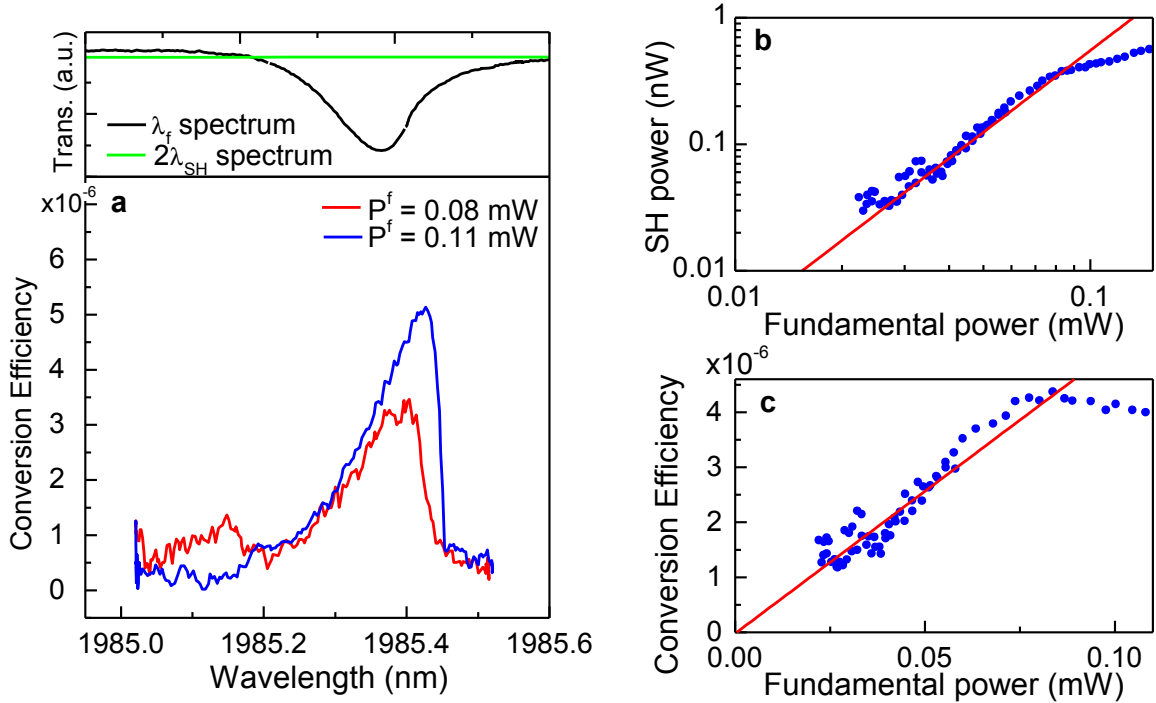

**Figure 5 Measure SHG spectra and dependence of SH conversion on fundamental pump power. a,** Microdisk SH conversion spectra (at two fundamental powers) as a function of fundamental pump wavelength. The top inset shows the low-power resonance spectra for the fundamental and SH (in units of $2\lambda_{SH}$) wavelengths. The SH resonance is centered at $2\lambda_{SH}$=1983.2 nm (off-scale). **b,** Logarithmic plot of generated second-harmonic vs. pump power at 1985.38-nm pump wavelength (blue dots). The fitted slope (red line) is 2.15±0.10, indicating quadratic growth of SH power with pump power. **c,** SHG conversion efficiency as function of 1985.38-nm pump power. The slope at low power indicates (5.2±0.4)×10$^{-5}$/mW normalized conversion efficiency, but the curve rolls off due to thermal distortion of the resonance.

2.2 nm, the theoretical η is 1.0×10$^{-3}$/mW[17]. The discrepancy between observed and theoretical η is due to uncertainty in the quality factors, especially at $\lambda_{SH}$. If the resonances of this microdisk were aligned ($|\lambda_f - 2\lambda_{SH}|=0$ nm), we expect 6.0×10$^{-2}$/mW. In comparison, SHG in a conventional, 5-mm-long, quasi-phasematched AlGaAs waveguide showed 2.3×10$^{-4}$/mW normalized conversion efficiency[24]. The GaAs microdisk is 1000 times smaller than the waveguide device and offers comparable normalized conversion efficiency.

The SH conversion in the GaAs microdisk is limited by thermal effects and higher-order nonlinearities. As seen in Fig. 5a, the generated SH spectrum is asymmetric even at fundamental powers as low as 80 µW; the asymmetry in the microdisk SHG spectrum mirrors that of the transmission spectrum at the fundamental wavelength. This type of power-dependent, lineshape distortion has been observed in other microphotonic devices[25–27] and is attributed to thermal and higher-order nonlinear effects. Since the fundamental wavelength is beyond the two-photon-absorption cut-off, linear absorption from surface or bulk defects is likely the primary cause of the lineshape distortion.

We have demonstrated, for the first time, $\bar{4}$-quasi-phasematched, second-harmonic generation in a GaAs microdisk cavity. $\bar{4}$-quasi-phasematching is a new QPM technique that relies on curved propagation geometries rather than engineered domain inversions. $\bar{4}$-QPM will enable second-order nonlinear optics in whispering-gallery-mode, nonlinear semiconductor resonators, which can be integrated in all-optical circuits. The normalized conversion efficiency is comparable to waveguide, quasi-phasematched GaAs devices and can be significantly improved by better tuning[28] to achieve doubly- rather than singly-resonant frequency conversion. The absolute conversion efficiency is limited by the onset of lineshape distortion caused by thermal and higher-order nonlinearities, which may be ameliorated by surface treatments[27]. Optical-frequency down-conversion in GaAs microdisk cavities can be used to generate entangled photon pairs. Using other $\bar{4}$-symmetry materials (such as GaP and ZnSe) would allow degenerate down-conversion using telecommunication wavelengths. Frequency conversion using microdisk cavities would be useful as miniature sources of infrared radiation for spectroscopy and sensing. Since the conversion depends on the quality factor of the optical cavity, it may be possible to directly sense the presence of absorbing species by monitoring conversion in the microdisk.

**Methods**

Molecular-beam epitaxy was used to make the sample, which consisted of a 160-nm-thick GaAs layer on top of a 1.5-$\mu$m-thick $Al_{0.75}Ga_{0.25}As$ sacrificial layer. The lateral dimensions of the GaAs microdisk were defined by photolithography and a hydrobromic acid etch. A hydrofluoric acid etch was then used to remove most of the $Al_{0.75}Ga_{0.25}As$ layer, leaving a narrow pedestal supporting the GaAs microdisk (see Fig.1). The pedestal had negligible effect on the low-radial-order optical modes used in the experiment.

The experiment setup is shown in Fig. 3. The probe beam was produced by a continuous-wave, 1900-to-2015-nm tunable, periodically poled $LiNbO_3$ (PPLN) optical parametric oscillator (OPO) (Lockheed Martin Aculight Argos), whose linewidth was less than 1 MHz. A long-pass filter blocked any leakage pump light from the OPO. To measure the resonances of the GaAs microdisks in the SH-wavelength range, a portion of the OPO light was doubled to 950-1007 nm in an oven-mounted PPLN crystal and then sent through a dichroic filter to reject the residual OPO light. The fundamental and SH beams were launched into two separate optical fibers, which were alternately connected to the fiber taper that addressed the GaAs microdisks. The powers of both beams were monitored using a 1% fiber tap coupler for the fundamental-wavelength beam and an uncoated BK7 pickoff plate placed near Brewster's angle for the SH-wavelength beam. A fiber polarization-controller was placed immediately before the fiber taper to control the polarization of the light at the microdisk. Light in the 1900-to-2015-nm-range was detected by an extended-range InGaAs detector. Light in the 950-to-1007-nm-range was detected by a Si detector preceded by dichroic mirror that strongly reflected any 1900-to-2015-nm light (to avoid possible two-photon absorption by the Si detector). All detected signals were sent to lock-in amplifiers referenced to a chopper at 205 Hz, and the lockin signal amplitudes were calibrated to an optical power meter. The OPO wavelength was constantly monitored by a wavemeter. The fiber taper and microdisk samples were mounted on two separate three-axis stages and imaged using a 50x microscope objective connected to a silicon camera. The temperature of the microdisks was stabilized at 31°C using a Peltier heater. The taper was fabricated by pulling a SMF28 fiber using motion-controlled stages while heating with a hydrogen-oxygen torch. The taper diameter was approximately 500 nm, where the fiber is single mode at both the fundamental and SH wavelengths with observed transmissions of 45% and 65%, respectively.

**References**

1. Armstrong, J. A., Bloembergen, N., Ducuing, J., & Pershan, P. S. Interactions between light waves in a nonlinear dielectric. *Phys. Rev.* **127**, 1918–1939 (1962).
2. Fejer, M. M., Magel, G. A., Jundt, D. H. & Byer, R. L. Quasi-phase-matched 2nd harmonic-generation - tuning and tolerances. *IEEE J. Quantum Electron.* **28**, 2631–2654 (1992).
3. Dumeige, Y. & Féron, P. Whispering-gallery-mode analysis of phase-matched doubly resonant second-harmonic generation. *Phys. Rev. A* **74**, 063804 (2006).
4. Yang, Z., Chak, P., Bristow, A. D., van Driel, H. M., Iyer, R., Aitchison, J. S., Smirl, A. L. & Sipe, J. E. Enhanced second-harmonic generation in AlGaAs microring resonators. *Opt. Lett.* **32**, 826–828 (2007).
5. Kuo, P. S., Fang, W. & Solomon, G. S. $\bar{4}$-quasi-phase-matched interactions in GaAs microdisk cavities. *Opt. Lett.* **34**, 3580–3582 (2009).
6. Yang, Z. & Sipe, J. E. Generating entangled photons via enhanced spontaneous parametric downconversion in AlGaAs microring resonators. *Opt. Lett.* **32**, 3296–3298 (2007).
7. Clemmen, S., Phan Huy, K., Bogaerts, W., Baets, R. G., Emplit, Ph. & Massar, S. Continuous wave photon pair generation in silicon-on-insulator waveguides and ring resonators. *Opt. Express* **17**, 16558–16570 (2009).
8. Van, V., Ibrahim, T. A., Absil, P. P., Johnson, F. G., Grover, R. & Ho, P.-T. Optical Signal Processing Using Nonlinear Semiconductor Microring Resonators. *IEEE J. Sel. Top. Quantum Electron.* **8**, 705–713 (2002).
9. Van, V., Ibrahim, T. A., Ritter, K., Absil, P. P., Johnson, F. G., Grover, R., Goldhar, J.& Ho, P.-T. All-Optical Nonlinear Switching in GaAs–AlGaAs Microring Resonators. *IEEE Photon. Tech. Lett.* **14**, 74–76 (2002).
10. Yariv, A., Xu, Y., Lee, R. K. & Scherer, A. Coupled-resonator optical waveguide: a proposal and analysis. *Opt. Lett.* **24**, 711–713 (1999).
11. Heebner, J. E., Boyd, R. W. & Park, Q-H. Slow light, induced dispersion, enhanced nonlinearity, and optical solitons in a resonator-array waveguide. *Phys. Rev. E* **65**, 036619 (2002).
12. Dumeige, Y. Quasi-phase-matching and second-harmonic generation enhancement in a semiconductor microresonator array using slow-light effects. *Phys. Rev. A* **83**, 045802 (2011).
13. Hafezi, M., Demler, E. A., Lukin, M. D. & Taylor, J. M. Robust optical delay lines with topological protection. *Nature Phys.* **7**, 907–912 (2011).

14. Maker, P. D., Terhune, R. W., Nisenoff, M. & Savage, C. M. Effects of dispersion and focusing on the production of optical harmonics. *Phys. Rev. Lett.* **8**, 21–22 (1962).
15. Angell, M. J., Emerson, R. M., Hoyt, J. L., Gibbons, J. F., Eyres, L. A., Bortz, M. L. & Fejer, M. M. Growth of alternating <100>/<111>-oriented II–VI regions for quasi-phasematched nonlinear optical devices on GaAs substrates. *Appl. Phys. Lett.* **64**, 3107–3109 (1994).
16. Ueno, Y., Ricci, V. & Stegeman, G. I. Second-order susceptibility of $Ga_{0.5}In_{0.5}P$ crystals at 1.5 μm and their feasibility for waveguide quasi-phase matching. *J. Opt. Soc. Am. B* **14**, 1428–1436 (1997).
17. Kuo, P. S. & Solomon, G. S. On- and off-resonance second-harmonic generation in GaAs microdisks. *Opt. Express* **19**, 16898–16918 (2011).
18. Ilchenko, V. S., Savchenkov, A. A., Matsko, A. B. & Maleki, L. Nonlinear Optics and Crystalline Whispering Gallery Mode Cavities. *Phys. Rev. Lett.* **92**, 043903 (2004).
19. Fürst, J. U., Strekalov, D.V., Elser, D., Lassen, M., Andersen, U. L., Marquardt, C. & Leuchs, G. Naturally phase-matched second-harmonic generation in a whispering-gallery-node resonator. *Phys. Rev. Lett.* **104**, 153901 (2010).
20. Rivoire, K., Lin, Z., Hatami, F., Masselink, W. T., & Vučković. Second harmonic generation in gallium phosphide photonic crystal nanocavities with ultralow continuous wave pump power. *Opt. Express* 17, 22609–22615 (2009).
21. Absil, P. P., Hryniewicz, J.V., Little, B. E., Cho, P. S., Wilson, R. A., Joneckis, L. G.& Ho, P.-T. Wavelength conversion in GaAs micro-ring resonators. *Opt. Lett.* **25**, 554–556 (2000).
22. Shoji, I., Kondo, T., Kitamoto, A., Shirane, M. & Ito, R. Absolute scale of second-order nonlinear-optical coefficients. *J. Opt. Soc. Am.B* **14**, 2268–2294 (1997).
23. Knight, J. C., Cheung, G., Jacques, F. & Birks, T. A. Phase-matched excitation of whispering-gallery-mode resonances by a fiber taper. *Opt. Lett.* **22**, 1129–1131 (1997).
24. Yu, X., Scaccabarozzi, L. & Harris, J. S. Efficient continuous wave second harmonic generation pumped at 1.55 μm in quasi-phase-matched AlGaAs waveguides. *Opt. Express* **13**, 10742–10748 (2005).
25. Barclay, P., Srinivasan, K. & Painter, O. Nonlinear response of silicon photonic crystal microresonators excited via an integrated waveguide and fiber taper. *Opt. Express* **13**, 801–820 (2005).
26. Borselli, M., Johnson, T. J. & Painter, O. Accurate measurement of scattering and absorption loss in microphotonic devices. *Opt. Lett.* **32**, 2954–2956 (2007).
27. Shankar, R., Bulu, I., Leijssen, R. & Loncar, M. Study of thermally-induced optical bistability and the role of surface treatments in Si-based mid-infrared photonic crystal cavities. *Opt. Express* **19**, 24828–24837 (2011).
28. Hennessy, K., Badolato, A., Tamboli, A., Petroff, P. M., Hu, E., Atatüre, M., Dreiser, J.& Imamoğlu, A. Tuning photonic crystal nanocavity modes by wet chemical digital etching. *Appl. Phys. Lett.* **87**, 021108 (2005).

**Acknowledgements**

We thank Angus Henderson for assistance with the optical parametric oscillator, and Kartik Srinivasan and Tim Thomay for helpful discussions.

**Author contributions**

P.K. and G.S. conceived the experiments; G.S. made the sample; P.K. processed the sample; P.K. performed the experiments and theoretical calculations, and analyzed the data; and P.K. and G.S. wrote the manuscript.

**Competing financial interests**

The authors declare no competing financial interests.